\documentclass[aps,prl,twocolumn,tightenlines,amsmath,amssymb,superscriptaddress]{revtex4-1}

\usepackage{graphicx}
\usepackage{amsmath}
\usepackage{amssymb}
\usepackage{epstopdf}
\usepackage{color}
\usepackage{lipsum}
\usepackage{ulem}
\usepackage[dvipsnames]{xcolor}
\usepackage[colorlinks=true]{hyperref} 
\usepackage[sort&compress]{natbib}

\DeclareGraphicsExtensions{.eps}

\newcommand{\ket}[1] {|#1 \rangle}

\newcommand{\gtwo}{{g}^{(2)}(0)}
\newcommand*{\rom}[1]{\expandafter\@slowromancap\romannumeral #1@}

\begin{document}

\title{{Bright Polarized Single-Photon Source Based on a Linear Dipole}}

	\author{S. E. Thomas$^{1}$, M. Billard$^2$,  N. Coste$^{1,2}$, S. C. Wein$^3$, Priya$^{1}$,  H. Ollivier$^{1}$, O. Krebs$^{1}$, L. Taza\"{i}rt$^1$,   A. Harouri$^1$, A. Lemaitre$^1$, I. Sagnes$^1$, C. Anton$^{1}$, L. Lanco$^{1,4}$, N. Somaschi$^2$, J. C. Loredo$^{*,1}$, and P. Senellart$^{+,}$}
	
	\affiliation{Centre for Nanosciences and Nanotechnology, CNRS, Universite Paris-Saclay, UMR 9001,
10 Boulevard Thomas Gobert, 91120, Palaiseau, France\\
        $^2$Quandela SAS, 10 Boulevard Thomas Gobert, 91120, Palaiseau, France\\
        $^3$Institute for Quantum Science and Technology and Department of Physics and Astronomy,
University of Calgary, Calgary, Alberta, Canada T2N 1N4 \\
		$^4$Universite Paris Diderot - Paris 7, 75205 Paris CEDEX 13, France\\
		$^*$ email: juan.loredo1@gmail.com\\
		$^+$ email: pascale.senellart-mardon@c2n.upsaclay.fr
		}

\begin{abstract}

{ Semiconductor quantum dots in cavities are promising single-photon sources. Here, we present a  path to deterministic operation, by harnessing the intrinsic linear dipole in a neutral quantum dot via phonon-assisted excitation. This enables emission of fully polarized single photons, with a measured degree of linear polarization up to 0.994 $\pm$ 0.007, and high population inversion -- 85\% as high as resonant excitation. We demonstrate a single-photon source with a polarized first lens brightness of 0.50 $\pm $ 0.01, a single-photon purity of 0.954 $\pm$ 0.001 and single-photon indistinguishability of 0.909 $\pm$ 0.004. }

\end{abstract}

\maketitle

The path towards an optimal single-photon source requires finding a scheme in which single photons are generated in a well-defined spatial and polarization mode, with near-unity efficiency, purity and indistinguishability. A device with these properties is highly sought-after for the advancement of quantum technologies such as secure long-distance quantum communication~\cite{Takemoto2015} and quantum computing~\cite{Slussarenko2019,Wang2020, Rudolph2017}. Several platforms towards optimised single-photon sources are being developed, including spontaneous parametric down-conversion (SPDC)~\cite{Ramelow:13,Weston:16,Kaneda:16} and four-wave mixing (FWM)~\cite{Spring:17,Francis-Jones:16}. These non-linear optical sources have an intrinsic efficiency limitation, and various multiplexing schemes are currently being explored to overcome this~\cite{Francis-Jones:16,Xiong2016,Joshi2018,Kaneda2019,Migdall2020}. Sources based on semiconductor quantum dots (QDs) in microcavities have recently shown their capability to deliver high purity, indistinguishable single photons with record  brightness~\cite{Somaschi2016,Ding2016,Senellart2017,He2019,Wang2019}. With an efficiency per photon more than one order of magnitude higher than sources based on frequency conversion, QD sources have already allowed a substantial scaling up of optical quantum computing~\cite{20photon}.

Despite this impressive progress, current QD sources are still far from the ideal deterministic performance that would provide a  single photon {in a pure quantum state} with unity probability. {This ambitious goal requires a scheme that maximises every single parameter controlling the source efficiency, including full inversion of the quantum dot transition, followed by the emission of a single photon with unity quantum purity, and perfect collection of the photon into a well-defined polarized optical mode.} Coherent control of a QD  has allowed generation of single photons in pure quantum states~\cite{He2013}, and near-unity population inversion has been obtained through more sophisticated techniques such as rapid adiabatic passage~\cite{Wu2011,Wei2014}. Unpolarized {collection} efficiencies of up to $78\%$ have been demonstrated with  micropillar cavities~\cite{Somaschi2016,Ding2016}. However, obtaining such high performance into a polarized mode remains challenging. So far, the most efficient sources have used resonant excitation of a charged exciton state, which inherently emits unpolarized photons. When using unpolarized cavities, only half of the single photons are collected through polarization filtering~\cite{Somaschi2016,Ding2016,Ollivier2020}. Very recently, polarized cavities were used to accelerate spontaneous emission into one linear polarization, providing nearly a factor of two gain in the source efficiency~\cite{Wang2019,Tomm2021}. Here, we propose a different path towards deterministic operation  -- making use of the linearly-polarized optical transitions of a neutral quantum dot, which arise due to its natural anisotropy, to directly generate and collect polarized single photons.

\begin{figure}[t]
		\centering
		\includegraphics[width=0.9\linewidth]{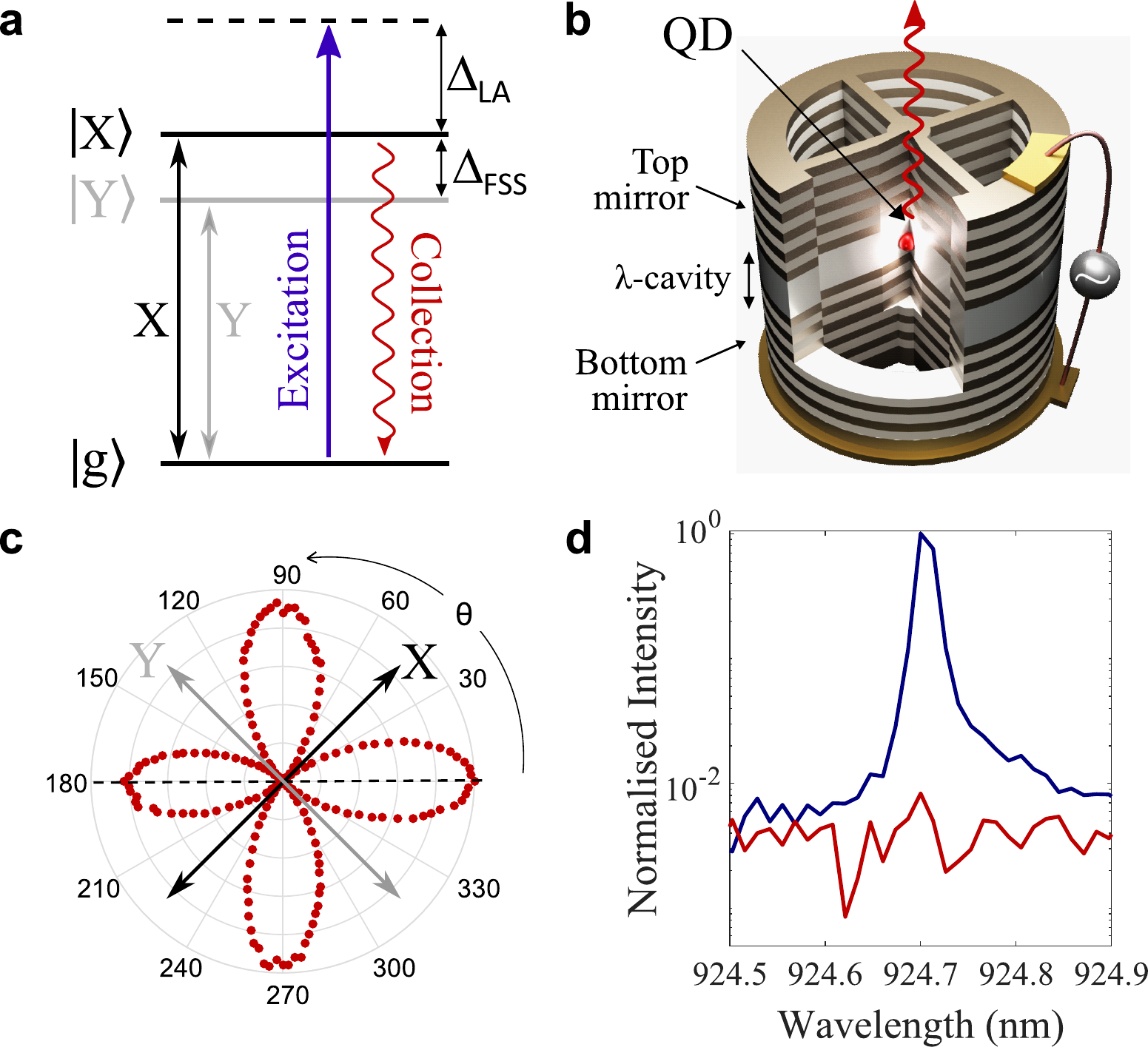}	
		\caption{(a) Energy level structure of a neutrally charged quantum dot, with exciton eigenstates $\ket{X}$ and $\ket{Y}$, separated in energy by $\Delta_\mathrm{FSS}$. The excitation pulse is blue-detuned from the transition by an energy of $\Delta_\mathrm{LA}$, or equivalently detuned in wavelength by $\Delta \lambda$. (b) { A schematic of the quantum dot deterministically embedded at the centre of an electrically-contacted micropillar cavity~\cite{Somaschi2016} } (c) Polar plot {of the cross-polarized emission intensity measured under phonon-assisted excitation as a function of the incident polarization angle}. (d) Spectrum of the emitted photons when exciting only the $X$ dipole of a neutral exciton and collecting in the parallel (blue) or orthogonal (red) basis. } 
	\label{fig:1}
	\end{figure}

 Neutral QDs have a three-level energy structure, as shown in Fig.~\ref{fig:1}(a) where two excitonic eigenstates, labelled $\ket{X}$ and $\ket{Y}$,  are separated by the fine structure splitting, $\Delta_\mathrm{FSS}$~\cite{Bayer2002}. The optical transition between the QD ground state $\ket{g}$ and one of the exciton states corresponds to a linearly-polarized dipole. While optical quantum technologies require polarized single photons, direct use of such a linear dipole has not been considered so far. Indeed, the generation of single photons with near-unity indistinguishability has only been reached under resonant excitation, a technique which is not applicable to a single linear dipole since the single photons have the same wavelength and polarization as the excitation laser and cannot be easily separated. 

To overcome this limitation, we propose the use of an off-resonant phonon-assisted excitation scheme,  which uses a slightly spectrally-detuned laser that can be easily separated from the single photons. This excitation scheme has recently been theoretically proposed to reach both high QD inversion probability and high quantum purity~\cite{Barth2016,Cosacchi2019,GustinHughes2020}, and relies on a detuned strong optical pulse that dresses the {ground and excited} states of the optical transition. During the pulse duration, the system relaxes between the dressed states through the emission of longitudinal-acoustic (LA) phonons. A strong occupation of the excited state is obtained following an adiabatic undressing of the dressed states during the switch-off of the excitation pulse. After the first observation of such phonon-assisted excitation~\cite{Quilter2015}, further theoretical studies predicted this excitation scheme should also provide near-unity single-photon purity, indistinguishability and occupation probability~\cite{Cosacchi2019,GustinHughes2020}. The combination of this phonon-assisted excitation scheme with the use of a QD linear dipole thus appears as a promising route towards deterministic polarized single photon sources.

We study LA-phonon-assisted excitation of a linear dipole in a neutral QD deterministically located at the center of an electrically connected micropillar cavity (Fig.~\ref{fig:1}(b))~\cite{Somaschi2016}. We study two samples (A and B) with a quality factor of around 4,000 (10,000) respectively. The cavities are almost circular and exhibit two linearly-polarized cavity modes typically separated by $\approx70~\mu$eV ($\approx30~\mu$eV), more than four times smaller than the cavity linewidth of $\approx 300~\mu$eV ({$\approx150~\mu$eV}). These devices have previously been shown to provide reproducible, high-performance single-photon generation under strictly resonant excitation,  where the excitation laser was separated from the emitted photons via cross-polarization. This reduced the polarized collection efficiency by a factor of two, and the polarized first lens brightness $\mathcal{B}_\mathrm{FL}$ -- the probability per pulse to collect a polarized single photon at the first objective lens above the micropillar cavity -- was limited to at best $\mathcal{B}_\mathrm{FL} \approx 25\%$~\cite{Somaschi2016,Ollivier2020}. 

In the present work, we implement LA-phonon-assisted excitation and set the excitation laser to be blue-detuned from the QD transition by approximately 0.6~nm, corresponding to a phonon energy of around 1~meV. The excitation pulse is derived from a 3~ps pulsed Ti-Sapphire laser centred at around 924.2~nm, which is shaped using a 4-f filtering system to obtain pulses with tunable temporal length from 12~ps to 22~ps and negligible spectral overlap with the QD emission at 924.8~nm. The single photon emission has a bandwidth of approximately 5~pm (10~pm) for Sample A (B), and is separated from the pump laser using three high-transmission 0.8~nm bandpass spectral filters. The samples are placed in a closed-cycle cryostat and cooled to around 8~K (7~K) for Sample A (B).  For more experimental details, see the {Supplemental} Material\cite{Supplemental}.

We can visualize the two linear dipoles of the neutral QD by varying the linear polarisation of the excitation laser, and measuring the single photon emission in the orthogonal polarization, as shown in Fig.~\ref{fig:1}(c). The intensity collected in cross polarization goes to zero when the polarization is aligned along one of the exciton linear dipoles and is maximum in between~\cite{Ollivier2020}. {We can set the polarization of the excitation laser to excite just one of the excitonic dipoles, and the system reduces to an effective two-level system, $\{ \ket{g}, \ket{X} \}$.} The measured spectra of the single photons when collecting parallel or orthogonal to the $X$ polarization direction, are shown in  Fig.~\ref{fig:1}(d), {evidencing strongly polarized emission}. By calculating the integrated intensity measured in parallel (crossed) polarization $I_{\parallel}$ ($I_{\perp}$), we find a degree of linear polarization, $D_\mathrm{LP}=\frac{I_{\parallel}-I_{\perp}}{I_{\parallel}+I_{\perp}}=0.994\pm0.007$ for device A1 on Sample A. {The same measurement was performed for two further different exciton-based devices on Sample A (devices A2 and A3) demonstrating linear polarization degree  $D_\mathrm{LP}$ of $0.981 \pm 0.004$ and $0.971 \pm 0.001$}. This polarization degree significantly exceeds what is accessible using polarized Purcell effect in asymmetric cavities, with maximal values of $0.92$ in optimized structures~\cite{Wang2019,Tomm2021} -- a limitation that arises from a compromise between reducing the cavity linewidth (to obtain large cavity splitting to linewidth ratios), and remaining in the weak coupling regime. With the current approach, there is no fundamental limit to the degree of polarization that could be reached. Even if the two linear dipoles present a slight non-orthogonality, which arises when the quantum dot shape and strain anisotropies do not coincide~\cite{Leger2007,Kowalik2008}, it is always possible in principle to excite just a single dipole by aligning the excitation light to be orthogonal to the other dipole axis. The values of $D_\mathrm{LP}$ that we measured are slightly less than the ideal value of one due to a slight imperfection in the precise alignment of the polarization axis inside the cavity.

	\begin{figure}
		\centering
		\includegraphics[width=.8\linewidth]{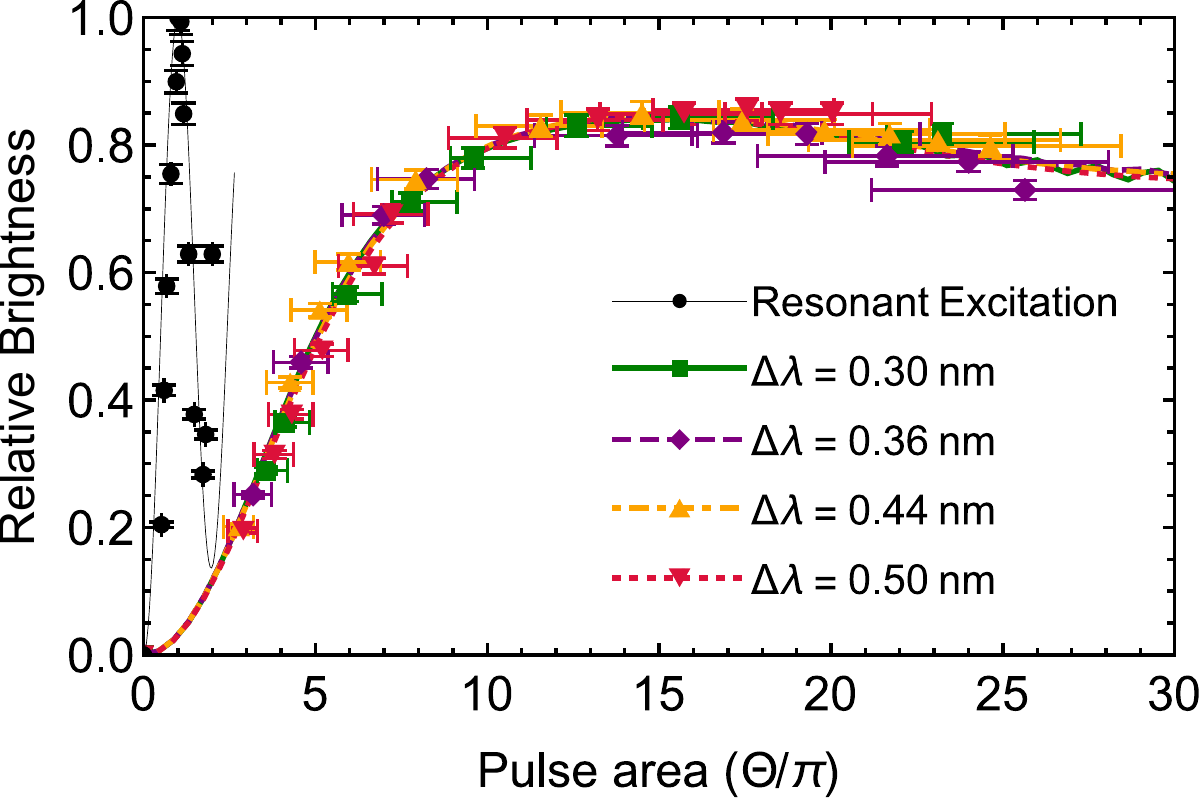}\vspace{-3mm}	
		\caption{ Measured intensity of single-photon emission from a charged exciton on Sample B collected in cross-polarization for resonant excitation (black) and LA-phonon-assisted excitation at various detunings,  $\Delta \lambda$, as a function of the pulse area experienced by the quantum dot inside the cavity. The brightness is normalised to the maximum intensity achieved under resonant excitation. The sample is cooled to {7~K} and the measured pulse duration is ($17 \pm 2$)~ps. The solid lines give the theoretical prediction based on the model in~\cite{GustinHughes2020} for a quasi-Gaussian pulse with FWHM duration of 16~ps, and the lines are mostly super-imposed for the detunings explored.}  
	\label{Figure2}
	\end{figure}

We now investigate the QD occupation probability, $p_\mathrm{QD}^\mathrm{LA}$, that can be obtained through LA-phonon-assisted excitation compared to the one achieved under resonant excitation $p_\mathrm{QD}^\mathrm{RF}$. To do so, we compare the single photon emission in a cross-polarization configuration, to be able to separate single photons from the laser for both excitation schemes. We study a charged exciton on Sample B, which corresponds to a four-level system that acts as an effective two-level system when collecting the emission in cross-polarization~\cite{Ollivier2020}, {in order to directly compare the two excitation schemes}. Fig.~\ref{Figure2} presents the single-photon counts detected under strictly resonant excitation (black) and using LA-phonon-assisted excitation (coloured) for a pulse duration of {$(17 \pm 2)$~ps} and various detunings from resonance as a function of the pulse area experienced by the quantum dot. The pulse area is proportional to the square root of the intra-cavity power, which is obtained by using the measured optical input power sent onto the cavity and the cavity transmission function at various detunings from the optical resonance. Higher input optical power is required to achieve the same intra-cavity pulse area for larger detunings. The error-bars on the intracavity pulse area come from  the uncertainty on the measured detunings. The resonant excitation data exhibits well-known Rabi oscillations reflecting the coherent control of the optical transition. Conversely, the LA-phonon-assisted excitation scheme does not show oscillations but a single rise and then slow decrease of the signal. The brightness of the single-photon emission is much less sensitive to both the laser power and detuning for LA-phonon-assisted excitation as compared to resonant excitation, which shows that this scheme is significantly more robust to experimental drifts and instabilities. The solid lines in Fig.~\ref{Figure2} show the theoretical predictions based on the model presented in~\cite{GustinHughes2020} for our experimental parameters, assuming a quasi-Gaussian pulse with a FWHM duration of 16~ps (see Supplementary Material for pulse shape). The theoretical model demonstrates good agreement with the experimental data. The measured occupation probability under LA-phonon-assisted excitation is as large as $0.85\pm 0.01$ compared to that reached under resonant excitation, demonstrating the high efficacy of this scheme. Such high occupation probability is also observed on device A3 as shown below and is on par with previous observations \cite{Ardelt14}. Theory indicates that the QD occupation probability could be brought even closer to unity using a lower temperature and developing appropriate temporal shaping of the excitation laser~\cite{Glassl2013}.

\begin{figure}
		\centering
		\includegraphics[width=.95\linewidth]{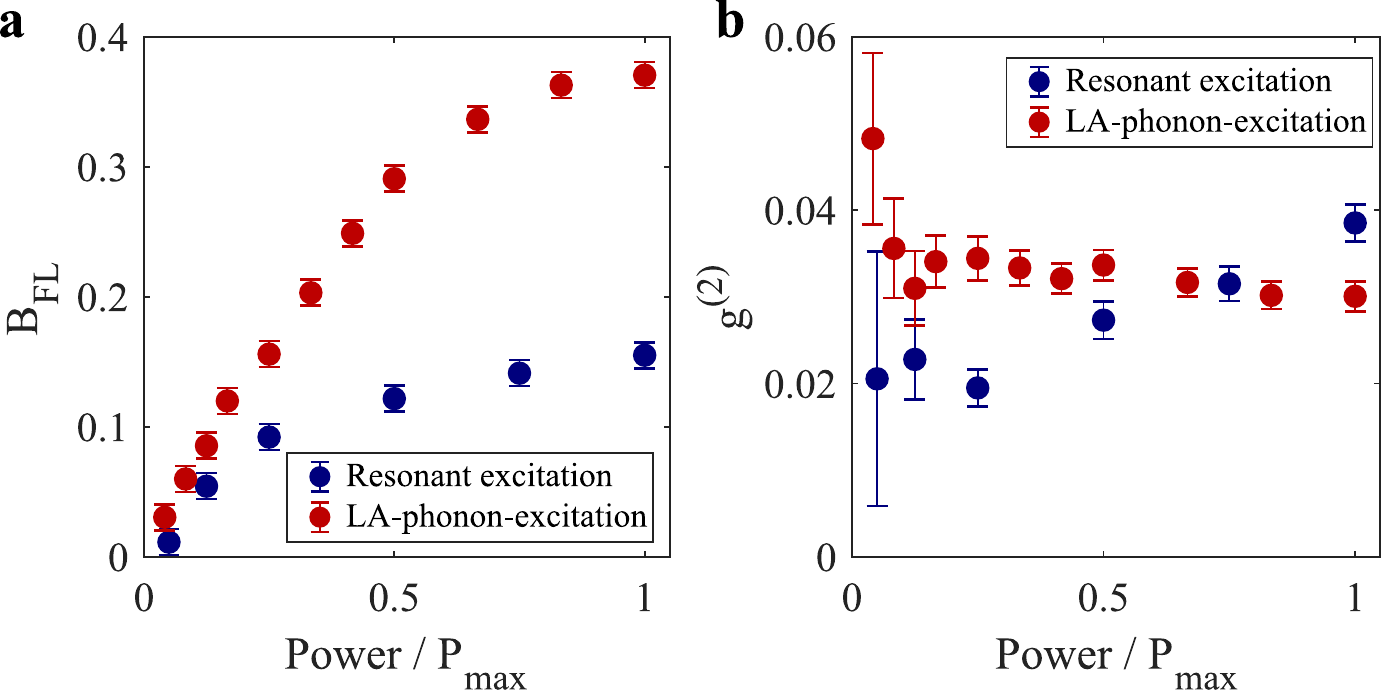}\vspace{-3mm}	
		\caption{ (a) First lens brightness and (b) $\gtwo$ as a function of the incident laser power for LA-phonon-assisted and resonant excitation, for Device A3 on Sample A. The power is scaled to the value required for maximum brightness for each excitation scheme, $P_\mathrm{max}$. The pulse duration is 13~ps in both cases, and the LA-phonon-assisted laser is blue-detuned from resonance by 0.85~nm. } 
		\label{Figure3}
	\end{figure}

We now focus on using a single linear dipole of a neutral exciton as a bright, polarized single-photon source. We investigate a neutral exciton (device A3) and measure the polarized brightness, single-photon purity and indistinguishability of the emitted photons. Fig.~\ref{Figure3}(a) shows the polarized first lens brightness and second-order autocorrelation, $\gtwo$, as a function of the laser power for resonant and LA-phonon-assisted excitation. In order to evaluate the first lens brightness, $\mathcal{B}_\mathrm{FL}$, we precisely calibrate all losses and the detector efficiency of our experimental setup (see Supplementary Material for the loss budget). For resonant excitation, the polarization of the excitation laser is aligned along one of the cavity axes and excites a combination of both $X$ and $Y$ optical transitions~\cite{Ollivier2020}. The brightness of the emission in cross-polarization depends on the angle between the dipole axes and the cavity axes, which for this particular device (Device 5 in Ref~\cite{Ollivier2020}) leads to a 15\% first lens brightness. For LA-phonon-assisted excitation the laser is aligned in polarization along the $X$ dipole axis, and blue-detuned by 0.85~nm with pulse duration of 13~ps. We can see that the polarized brightness reaches a value of 37\% for these parameters -- already more than a factor of two brighter than for resonant excitation. 

Fig.~\ref{Figure3}(b) shows the second-order autocorrelation, $\gtwo$ as a function of power for the two excitation schemes.  At maximum brightness we measure very similar single-photon purity for LA-phonon-assisted excitation ($\gtwo = 0.030 \pm 0.002$) and for resonant excitation ($\gtwo = 0.039 \pm 0.002$). We further note that a slight reduction of the $\gtwo$ is observed when increasing the excitation power for the phonon-assisted excitation scheme, in accordance with theoretical prediction~\cite{Cosacchi2019} and in contrast to the case of resonant excitation. Preliminary theoretical study indicates that the minimal reachable $\gtwo$ value strongly depends on the precise temporal shape of the laser and could be significantly reduced.

We also measure the Hong-Ou-Mandel interference visibility, $V_\mathrm{HOM}$, at maximum brightness for LA-phonon-assisted excitation. We measure $V_\mathrm{HOM} = 0.851 \pm 0.002 $, from which we extract the single-photon mean wavepacket overlap~\cite{g2HOM} of $M_\mathrm{s} =  (V_\mathrm{HOM} + \gtwo)/ (1 - \gtwo) = 0.911 \pm 0.003 $. For resonant excitation we measured $M_\mathrm{s} = 0.915 \pm 0.001$, demonstrating that the indistinguishability of the emitted photons under LA-phonon-assisted excitation is the same as the state-of-the-art values obtained under resonant excitation. We note that a high indistinguishability of around 84 \% had previously been demonstrated using phonon-assisted excitation of a QD in bulk~\cite{Reindl2019}. Our study shows that the acceleration of spontaneous emission needed to obtain bright sources does not prevent high values of indistinguishability to be reached under phonon-assisted excitation. In fact the use of a cavity actually allows further improvement of the indistinguishability via suppression of the phonon sideband.

\begin{figure}
		\centering
		\includegraphics[width=.95\linewidth]{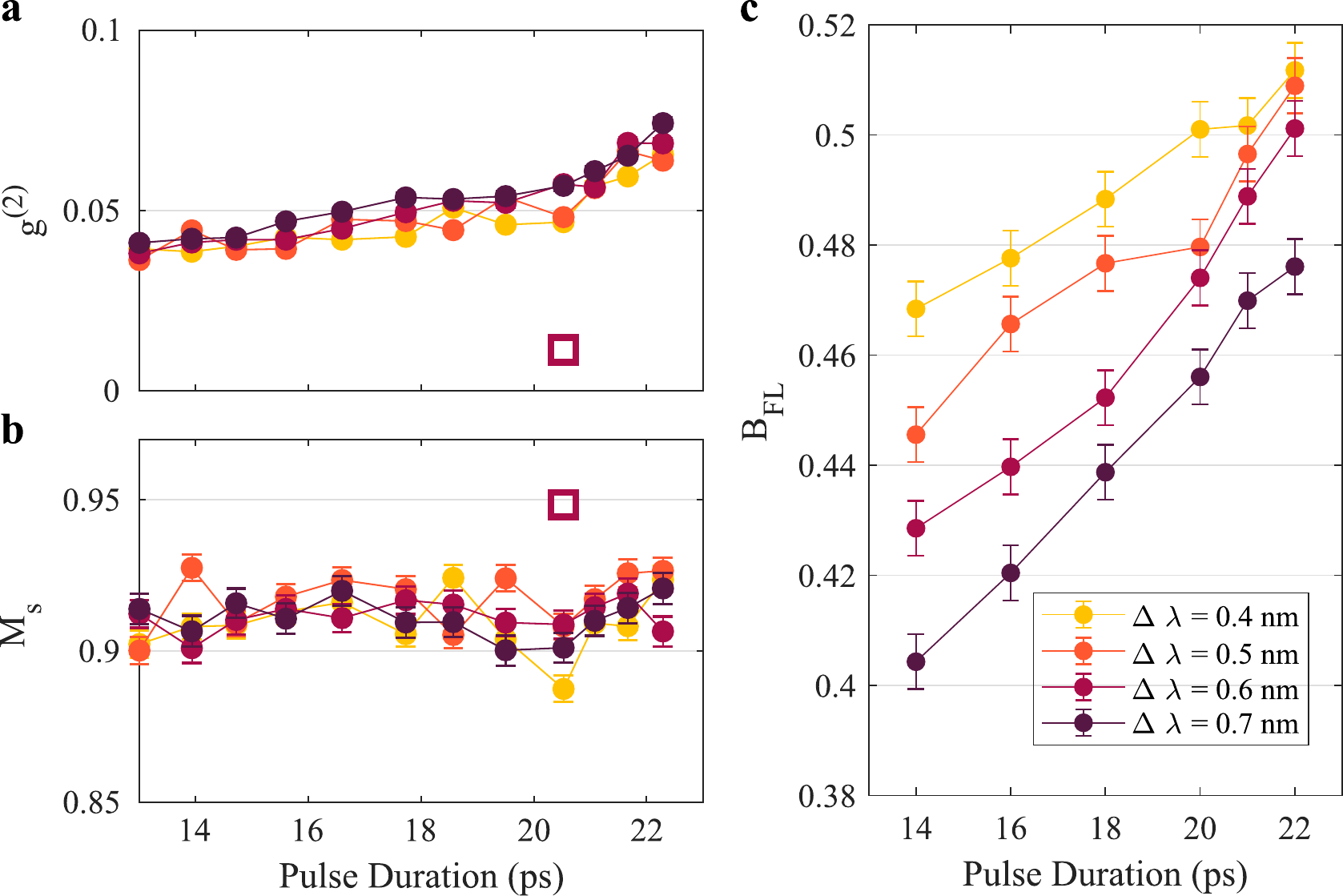}\vspace{-3mm}	
		\caption{ (a) The second-order autocorrelation, $\gtwo$, (b) single-photon indistinguishability, $M_\mathrm{s}$, and (c) the first lens brightness,  $\mathcal{B}_\mathrm{FL}$, as a function of the pulse duration and detuning $\Delta \lambda$ for Device A3 on Sample A. The open square points in (a) and (b) correspond to the values measured with an additional 10~pm bandwidth etalon filtering the collected photons.} 
	\label{Figure4}
	\end{figure}

To find the optimal operational conditions for a bright single-photon source, we measure the single-photon purity, indistinguishability and brightness of device A3 as a function of the detuning, $\Delta \lambda$, and  pulse duration of the excitation laser, $\tau$. For each set of parameters, we adjust the excitation power to reach maximum brightness. The $\gtwo$ {remains mostly unchanged up to 18 ps pulse duration and} increases for longer pulses due to a higher probability of re-excitation, as shown in Fig.~\ref{Figure4}(a). The single-photon indistinguishability is approximately constant at around 0.92 for all the explored parameters, as shown in Fig.~\ref{Figure4}(b). For one particular set of parameters we also measured the single-photon purity and indistinguishability with an additional 10~pm bandwidth etalon filtering the collected photons, as shown by the open squares in Fig.~\ref{Figure4}(a-b). For a detuning of 0.6~nm and a pulse duration of 20.5~ps we measured $\gtwo = 0.011 \pm 0.001 $ and $M_\mathrm{s} = 0.948 \pm 0.001$. This additional spectral filtering improves the $\gtwo$ by suppressing any spurious laser photons as well as any extra spectrally-broader photons due to re-excitation~\cite{g2HOM}. It also reduces the residual phonon-sideband emission and hence improves the indistinguishability. However, {the etalon used in the present experimental implementation shows limited peak transmission.} With improved, efficient spectral filtering it will be possible to reach significantly better values of single-photon purity and indistinguishability whilst maintaining the high brightness. 

Fig.~\ref{Figure4}(c) shows that higher brightness is achieved for smaller detunings which agrees with the theoretical prediction that the phonon-assisted excitation process becomes ineffective for larger detunings, and the optimal detuning will depend on the specific quantum dot and phonon coupling parameters~\cite{GustinHughes2020}. The brightness also increases for longer pulse durations, but with a slight degradation in the single-photon purity.  For a detuning of $\Delta \lambda = 0.4$~nm and a pulse duration of $\tau = 19.5$~ps, we obtain a first-lens brightness of $B_\mathrm{FL} = 0.50 \pm 0.01$, a single-photon purity of $\mathcal{P} = 1 - \gtwo = 0.954 \pm 0.001 $ and a single-photon indistinguishability of $M_\mathrm{s} = 0.909 \pm 0.004$. This is more than a factor of three higher than the brightness in resonant excitation for this device whilst maintaining state-of-the-art performance. Considering the extraction efficiency of our cavities $\eta_\mathrm{ext} \approx 0.65$, and the theoretically expected  ${p_\mathrm{QD}^\mathrm{LA}} \approx 0.85$, the maximum expected first-lens brightness is $\mathcal{B}_\mathrm{FL} = \eta_\mathrm{ext} p_\mathrm{QD} \approx 0.55$, very close to our experimental observation. This value could be improved using cavities with higher extraction efficiencies of $\eta_\mathrm{ext} = 0.80 $~\cite{Gazzano2013}, and by improving the QD occupation probability to near-unity via pulse shaping techniques~\cite{Glassl2013}. In our current experimental implementation, the $50\%$ first lens brightness corresponds to a detected count rate of 6~MHz using a 69\% efficient single photon detector and a laser repetition rate of 81~MHz. This value is limited by the low optical transmission of 17\% of our experimental set-up which could be considerably improved with optimized spectral filtering techniques such as custom fibre Bragg gratings which could filter the excitation laser from the single photons with an efficiency of over 90\%, and by direct coupling of the QD to an optical fibre with efficiencies of up to 90\%~\cite{Snijders2018}.

In conclusion, we have reported on a new approach to bring QD-based single photon sources closer to deterministic operation, by perfoming off-resonant phonon-assisted excitation of a linearly-polarized dipole of a neutral QD. We demonstrated that phonon-assisted excitation can enable  indistinguishabilities in the $90-95\%$ range, at the same level as  resonant excitation,  high occupation probability, and  a significant increase in source brightness. As with resonant excitation, the indistinguishability reported here is limited by the moderate quality factor of our cavity. Near unity values should be reached  using cavities with stronger Purcell effects as for resonant excitation~\cite{Somaschi2016}, which is key for practical quantum technologies. The use of phonon-assisted excitation also enables a significant gain in stability and robustness, since the scheme is resilient to drifts in QD-laser detuning and excitation power. All these features are of great importance for practical applications in quantum technologies. Finally we note that the use of phonon-assisted excitation in unpolarized cavities gives the possibility of exciting and collecting the photons in all polarization directions, which is critical for the generation of photonic cluster states using charged excitons~\cite{Lindner2009}, and this demonstration of indistinguishable single photons using this scheme is an important first step. 

\vspace{5mm}

\noindent \textbf{Acknowledgements.} The authors would like to thank Andrew White and Marcelo Pereira de Almeida for experimental help. This work was partially supported by the ERC PoC PhoW,  the IAD-ANR support ASTRID program Projet ANR-18-ASTR-0024 LIGHT, the QuantERA ERA-NET Cofund in Quantum Technologies project HIPHOP, the EU Horizon2020 FET OPEN project QLUSTER (Grant ID 862035), the EU Horizon2020 FET OPEN project PHOQUSING (Grant ID 899544), and the French RENATECH network, a public grant overseen by the French National Research Agency (ANR) as part of the "Investissements d’Avenir" programme (Labex NanoSaclay, reference: ANR-10-LABX-0035). J.C.L. and C.A. acknowledge support from Marie Skłodowska-Curie Individual Fellowships SMUPHOS and SQUAPH, respectively. S.C.W. acknowledges support from NSERC (the Natural Sciences and Engineering Research Council), AITF (Alberta Innovates Technology Futures), and the SPIE Education Scholarships program. H.O. and N.C. acknowledge support from the Paris Ile-de-France Région in the framework of DIM SIRTEQ.

\bibliography{la-sps}
\bibliographystyle{naturemag}

\clearpage
\onecolumngrid

	\appendix
	\setcounter{figure}{0} \renewcommand{\thefigure}{S.\arabic{figure}}
	\setcounter{equation}{0} 
	\renewcommand{\theequation}{S.\arabic{equation}}
	\setcounter{table}{0} 
	\renewcommand{\thetable}{S.\arabic{table}}

{\begin{center}
    \large{Supplementary Material}
\end{center}}

\section{Experimental Set-up}

To excite the QD via LA-assisted excitation we require a pulse that is blue-detuned from the QD resonance by $0.3 - 0.85$~nm. We start with a 3~ps pulsed Ti-Sapphire laser with a central wavelength around $924~$nm and a $81~$MHz repetition rate. The laser spectrum is shaped using a 4-f filtering system to obtain pulses with tunable temporal length from 3~ps to 22~ps, or equivalently a linewidth from 0.4~nm down to 0.06~nm. This has negligible spectral overlap with the QD emission which is centred at 924.82~nm and has a linewidth of approximately 10~pm. The co-linearly polarized single photons are collected with a non-polarizing beam-splitter transmitting 90\% of the light. Three high-transmission ($\sim{95}\%$ each), $0.8~$nm FWHM, high-extinction band-pass filters (Alluxa) separate the single photons from the excitation laser. A half and quarter waveplate are used to precisely align the polarization along the exciton axis ($X$ or $Y$) inside the cavity. Superconducting nanowire single photon detectors (SNSPDs) with 25-30~ps timing jitter and an efficiency of {75\%} at low count rates, and 69\% at the maximum measured count rate, are used to record the time dependence of the emission and the second order intensity correlations.

\begin{figure}[h]
	\centering
	\includegraphics[width=.3\textwidth]{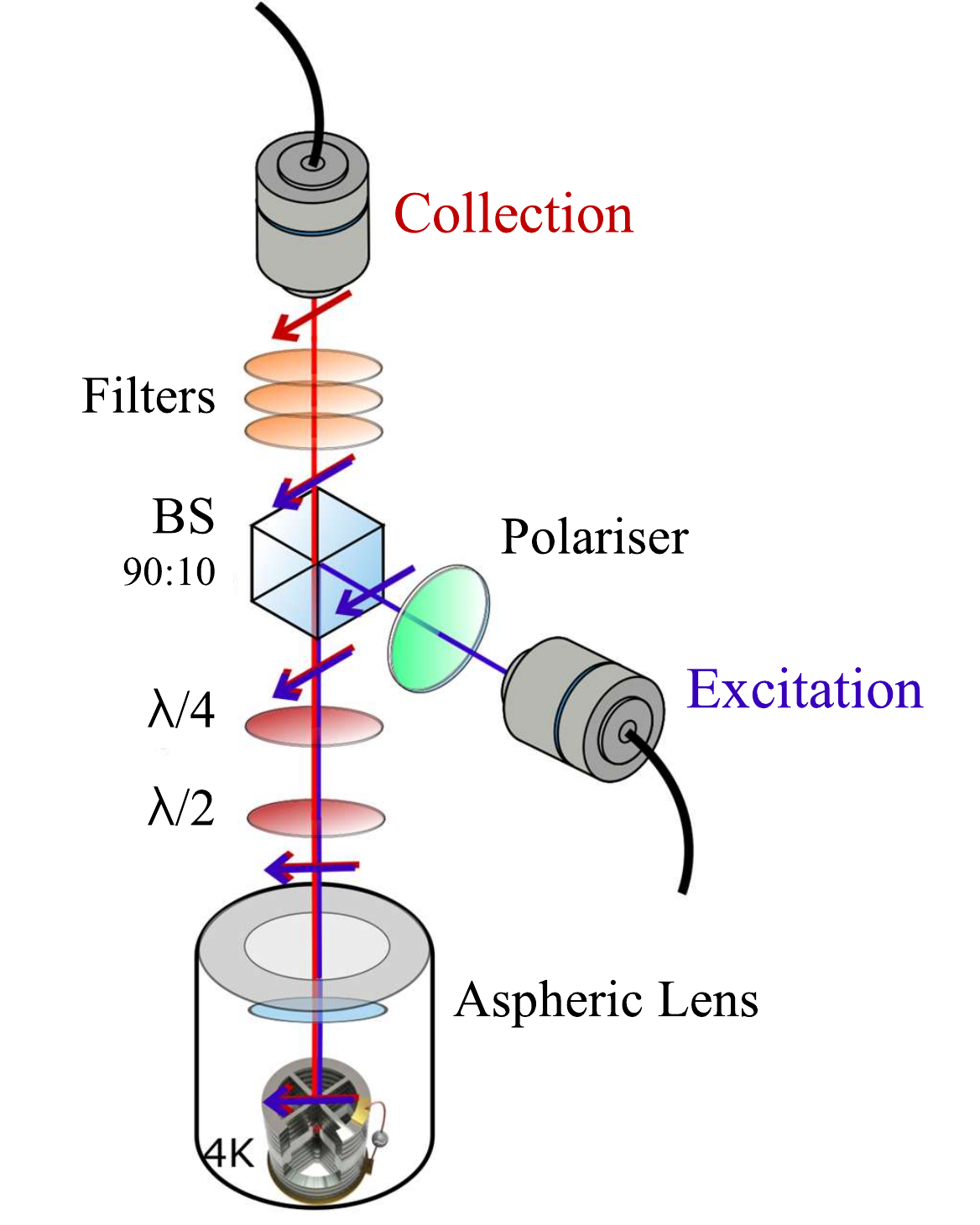}\vspace{-3mm}	
	\caption{ {The experimental set-up for phonon-assisted excitation.} }
	\label{FigureS1_Setup}
\end{figure}

\section{Loss Budget}

We measure the transmission of the optical components of our set-up using a continuous wave laser centered at the quantum dot emission wavelength, 924.82~nm, and details are given in Table~\ref{Table2}. To measure the transmission of the window of the cryostation and the objective lens, we measure the reflection of the beam off the flat diode surface of the sample, and assume that the reflectivity of the surface is 100\% in order to give an upper bound on the transmission of the optical elements. The light passes through the cryostation window, the objective lens and two waveplates twice, and we give the single-pass transmission of these combined elements in Table~\ref{Table2}. 

The transmission of the filters is less than the maximum of 95\% per filter, since we angle-tune the filters to fully suppress the excitation laser. The filters were aligned here for an excitation laser with a detuning of $\Delta \lambda = 0.8$~nm and a pulse duration of $\tau = 20$~ps. In order to calculate the first lens brightness at a different detuning or pulse duration, the filters are aligned to suppress the laser and then the relative transmission efficiency is measured with respect to this known point. 

\begin{figure}[h]
	\centering
	\includegraphics[width=.75\textwidth]{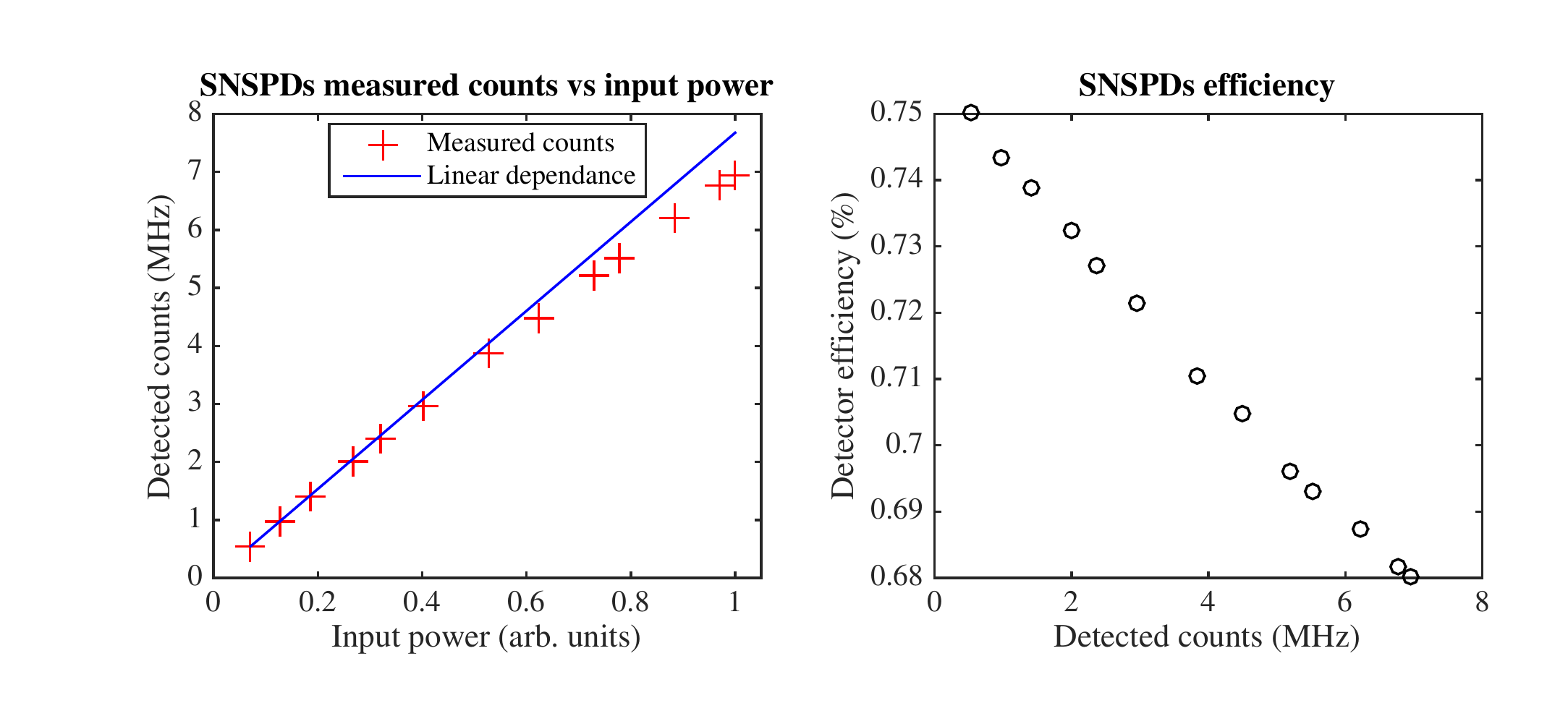}\vspace{-3mm}	
	\caption{ Left: Detected counts on an SNSPD as a function of the input power. Right: Detection efficiency of SNSPD as a function of the detected count rate.    }
	\label{FigureS5}
\end{figure}

The detector efficiency of the SNSPDs at a fixed bias current depends on the count rate because of the dead time of the detectors~\cite{Esmaeil2017}. We measured the detection efficiency using an attenuated laser pulse with a repetition rate of 81~MHz. The detection efficiency for a low count rate of 500~kHz was found to be 75\%. However, when the count rate increases to 6~MHz the detection efficiency decreases to 69\%, as shown in Figure~\ref{FigureS5}. We note that this may not be the optimal detection efficiency for these parameters as we did not adjust the bias current of the SNSPDs. 

The total efficiency of the set-up for a detuning of $\Delta \lambda = 0.8$~nm, a pulse duration of $\tau = 20$~ps, and a detected count rate of $R_\mathrm{det} = 6$~MHz is $\eta = 0.17 \pm 0.01$. The repetition rate of the laser $R_\mathrm{L} = 81$~MHz, and therefore this corresponds to a first lens brightness of $\mathcal{B}_\mathrm{FL} = R_\mathrm{det}/(R_\mathrm{L}\times \eta) = 0.44$ for these parameters. 

{The set-up losses could be significantly reduced by coupling the light from the micropillar cavity directly into an optical fibre~\cite{Snijders2018}. Furthermore, the bandpass filters could be replaced with custom fibre Bragg gratings which would allow for suppression of the excitation laser with significantly higher transmission of the single photons. }

\begin{table}[h]
\begin{center}
\begin{tabular}{|c|c|}
\hline
     Element & Transmission  \\
     \hline
     Cryostat Window + Objective Lens + QWP + HWP & 0.80 $\pm$ 0.02  \\
     Beam splitter & $0.92 \pm 0.01$ \\
     5 Mirrors  & $0.95 \pm 0.01$ \\
     Telescope & $0.98 \pm 0.01$\\
     2 Waveplates & $0.95 \pm 0.01 $\\
     3 Bandpass filters & $0.64 \pm 0.02$ \\
     Fibre coupling efficiency & $0.60 \pm 0.02$ \\
     Detection efficiency (low count rate) & $0.75 \pm 0.03$\\
     Detection efficiency (6 MHz detected rate) & $0.69 \pm 0.03$ \\
     \hline
\end{tabular}
\end{center}
\caption{\textbf{Loss Budget}. The transmission of optical elements in the set-up, and efficiency of the superconducting nanowire detectors \label{Table2} }
\end{table}

\section{Pulse Shape for Theoretical Model }

A detailed theoretical model for estimating the efficiency of LA-phonon-assisted excitation is given in~\cite{GustinHughes2020}. We apply this model with the relevant experimental parameters in order to understand how the brightness of the polarized single photon source depends on the detuning and pulse duration of the excitation pulse. 

\begin{figure}[h]
	\centering
	\includegraphics[width=.6\textwidth]{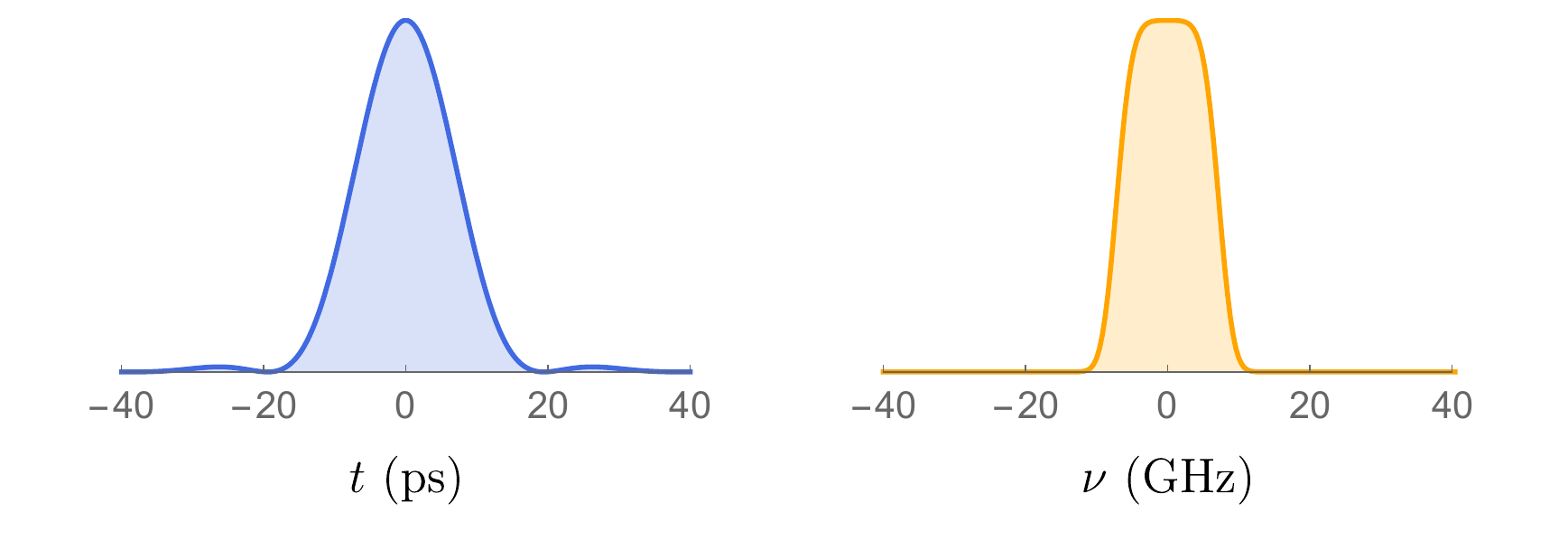}\vspace{-3mm}	
	\caption{Temporal profile (left) and spectrum (right) of the pulse used in the theoretical model for Figure 2 in the main text. }
	\label{FigureSPulseShape}
\end{figure}

{For the theoretical prediction presented in Figure 2 in the main text we use the pulse shape shown in Figure~\ref{FigureSPulseShape}. The spectrum is relatively close to a top hat due to the slit used in the 4f shaping line. The longer time components in the temporal pulse shape cause re-excitation of the quantum dot, and optimised shaping of the pulse would enable significantly lower values of $\gtwo$ to be reached. }

\end{document}